\documentclass[twocolumn]{aastex63}
\usepackage{amsmath}
\usepackage{amssymb}

\definecolor{purple}{rgb}{0.5,0,0.5}
\definecolor{darkgreen}{rgb}{0.1,0.6,0.1}
\definecolor{orange}{rgb}{1,0.6,0}

\submitjournal{ApJ}
\shorttitle{Magnetic Archaeology}
\shortauthors{Jermyn et al.}

\begin{document}

\title{Magnetic Archaeology of Early-Type Stellar Dynamos}

\correspondingauthor{Adam S. Jermyn}
\email{adamjermyn@gmail.com}

\author[0000-0001-5048-9973]{Adam S. Jermyn}
\affiliation{Center for Computational Astrophysics, Flatiron Institute, New York, NY 10010, USA }

\author[0000-0002-8171-8596]{Matteo Cantiello}
\affiliation{Center for Computational Astrophysics, Flatiron Institute, New York, NY 10010, USA }
\affiliation{Department of Astrophysical Sciences, Princeton University, Princeton, NJ 08544, USA}

\begin{abstract}

Early-type stars show a bimodal distribution of magnetic field strengths, with some showing very strong fields ($\ga 1\,\mathrm{kG}$) and others very weak fields ($\la 10\,\mathrm{G}$).
Recently, we proposed that this reflects the processing or lackthereof of  fossil fields by subsurface convection zones.
Stars with weak fossil fields process these at the surface into even weaker dynamo-generated fields, while in stars with stronger fossil fields magnetism inhibits convection, allowing the fossil field to remain as-is. 
We now expand on this theory and explore the time-scales involved in the evolution of near-surface magnetic fields.
We find that mass loss strips near-surface regions faster than magnetic fields can diffuse through them.
As a result, observations of surface magnetism directly probe the frozen-in remains of the convective dynamo.
This explains the slow evolution of magnetism in stars with very weak fields: these dynamo-\emph{generated} magnetic fields evolve on the time-scale of the mass loss, not that of the dynamo.

\end{abstract}


\section{Introduction}\label{sec:intro}

Massive stars play host to some of the most unusual stellar phenomena, including rapid mass loss~\citep{Smith:2014}, time-varying UV emission~\citep{2014MNRAS.441..910R}, strong magnetic fields~\citep{2013MNRAS.433.2497S}, stochastic low-frequency variability~\citep{2020FrASS...7...70B}, rotational mixing~\citep{doi:10.1146/annurev.astro.38.1.143}, opacity-driven pulsations~\citep{1993MNRAS.262..204D}, and massive outbursts~\citep{1994PASP..106.1025H}.
In part because of this their structure and evolution remain uncertain, with major open questions remaining as to the cause and extent of interior mixing~\citep{2021NatAs.tmp...80P}, their late stage evolution~\citep{2020svos.conf..223E}, and the compact remnants they leave behind.

Magnetic fields are central to understanding many of these phenomena.
Magnetism interacts strongly with stellar winds~\citep{2005ARA&A..43..103Z} and X-ray/UV emission~\citep{2010ARA&A..48..241B}, is a key component of many mixing processes in stars~\citep[e.g.][]{2019ApJ...879...60G,2019MNRAS.485.3661F}, and is known to affect with internal gravity waves and stellar pulsations~\citep{2015Sci...350..423F,2018A&A...616A.148B}.
Moreover magnetism interacts with convection in a number of ways~\citep{2014ARA&A..52..251C,Augustson_2016}, which may play a crucial role in setting the long-run evolution of stars.

Unfortunately stellar magnetism is challenging to understand theoretically, both because it frequently arises from complex non-linear processes such as dynamos~\citep{Augustson_2016} and because the relevant time-scales span many orders of magnitude, from those of stellar evolutionary and magnetic diffusion to those of the much more rapid Alfv\'enic adjustments~\citep{2011A&A...534A.140C,2017RSOS....460271B} and turbulent dynamos.
Our handle on magnetism therefore comes primarily from observations, which directly probe only the surface layers of a star, though asteroseismology can in some cases yield insights into magnetism in deeper regions~\citep{2015Sci...350..423F}.

Here we explore the time-scales and evolution of near-surface magnetic fields in O/B/A stars.
We begin in Section~\ref{sec:times} by examining the relevant time-scales for magnetic diffusion, mass-loss, and convection.
We find that the convective turnover time is always the shortest of these, followed by mass-loss and then magnetic diffusion.
This ordering motivates a story which we describe in Section~\ref{sec:story}, whereby mass-loss causes subsurface convection zones to move steadily deeper in a star with time, trailing a frozen-in dynamo-generated magnetic field behind them.
We expect then that photometric and spectropolarimetry observations directly probe the frozen remains of a convective dynamo field, and discuss the possibilities for connecting this theory with observations.
We examine the limitations and challenges for our theory in Section~\ref{sec:limits}, and finally conclude with a summary of our findings in Section~\ref{sec:discussion}. 
\section{Time-Scales}\label{sec:times}

We calculated stellar evolutionary tracks for stars ranging from $2-60 M_\odot$ using revision 15140 of the Modules for Experiments in Stellar Astrophysics
\citep[MESA][]{Paxton2011, Paxton2013, Paxton2015, Paxton2018, Paxton2019} software instrument.
Details on the MESA microphysics inputs are provided in Appendix~\ref{appen:mesa}.
From these models we can extract several time-scales, which we discuss below.

\subsection{Magnetic Diffusion}

\begin{figure*}
\centering
\includegraphics[width=0.98\textwidth]{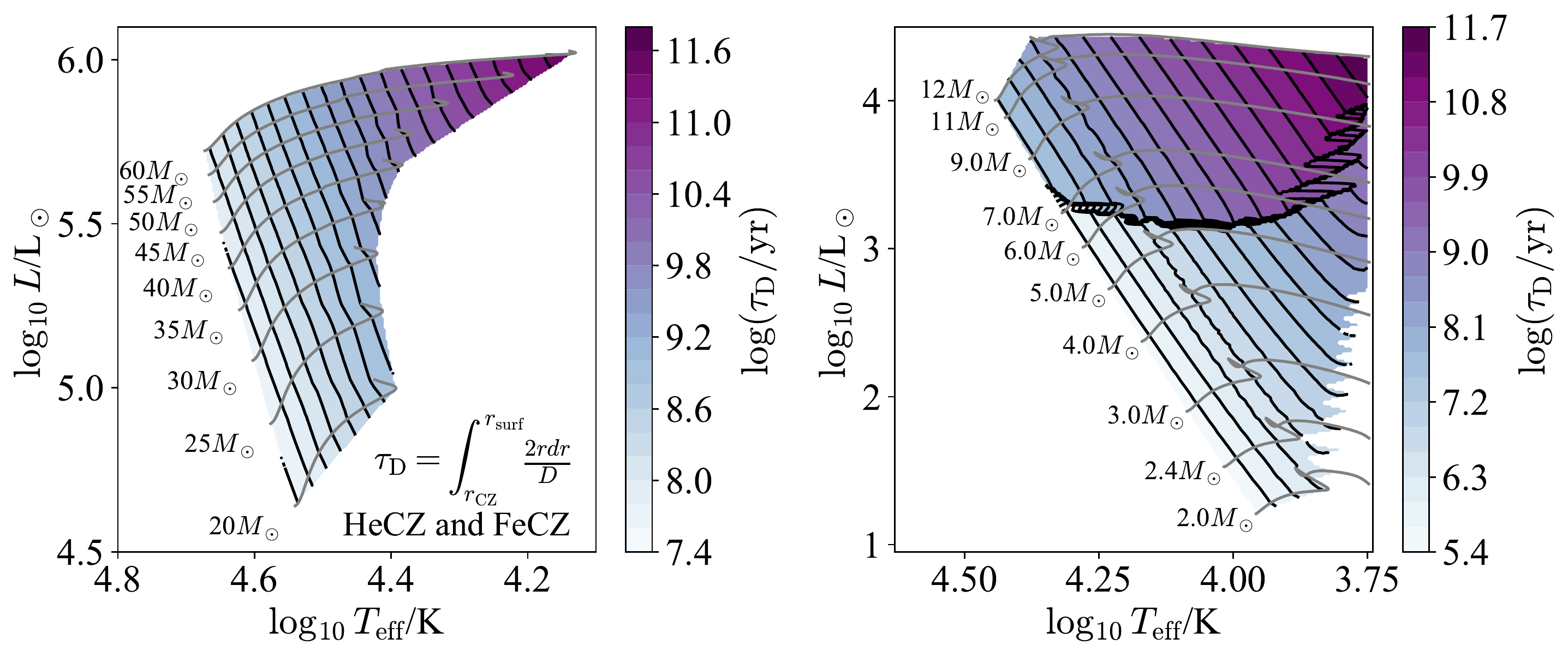}
\caption{The magnetic diffusion time-scale given by equation~\eqref{eq:tauD} is shown on a Hertzsprung-Russel diagram in terms of $\log T_{\rm eff}$ and $\log L$ for stellar models ranging from $2-12 M_\odot$ with $Z=0.02$. The sharp change at $\log L/L_\odot \approx 3.2$ is due to the appearance of the FeCZ. At low $\log T_{\rm eff}$ and low $\log L$ we have omitted regions where a vigorous H convection merges with the subsurface convection zones, and we do not report values calculated for the H convection zone in this diagram. Note that the two panels use different color scales.}
\label{fig:tauD}
\end{figure*}

The first time-scale is that of magnetic diffusion in the region above the subsurface convection zone, given by
\begin{align}
	\tau_{\rm D} \equiv \int_{r_{\rm CZ}}^{r_{\rm surf}} \frac{2 r dr}{D},
	\label{eq:tauD}
\end{align}
where $r_{\rm CZ}$ is the radius at the top of the subsurface convection zone, $r_{\rm surf}$ is the radius of the surface of the stellar model, near the photosphere, $D = \eta c^2 / 4 \pi$ is the magnetic diffusivity, $c$ is the speed of light, and $\eta$ is the Spitzer resistivity~\citep{1950PhRv...80..230C}, given by
\begin{align}
	\eta = \frac{4 \bar{Z} e^2  }{3 k_B T}\sqrt{\frac{2\pi m_e}{k_B T}}\ln \left(\frac{4}{3}\pi \lambda_D^3 n_e\right).
\end{align}
Here $\bar{Z}$ is the mean ion charge, $e$ is the electron charge, $m_e$ is the electron mass, $k_B$ is the Boltzmann constant, $T$ is the temperature, $n_e$ is the free electron density, $\lambda_D = \sqrt{k_B T/4\pi e^2 \bar{Z^2}}$ is the Debye wavelength, and $\bar{Z^2}$ is the mean square ion charge.

Figure~\ref{fig:tauD} shows this time-scale for the convection zone with the greatest turbulent kinetic energy on a Hertzsprung-Russel diagram in terms of $\log T_{\rm eff}$ and $\log L$.
In what follows we always examine just the dominant convection zone in this sense, and show results on the same Hertzsprung-Russel diagram.

For less massive stars ($M \la 7 M_\odot$) the dominant convection zone is the HeCZ, which lives very near the stellar surface, so this time-scale is short compared with the main-sequence lifetime of the star.
As a result magnetic fields are not frozen in on the time-scales of stellar evolution.
For more massive stars the strongest convection zone is the FeCZ.
Because the top of the FeCZ is deeper than that of the HeCZ the diffusion time-scale is longer, and in general exceeds the main-sequence lifetime of the star.
In general then there is a separation between O stars, for which the magnetic field is effectively frozen in on the stellar lifetime, and A/B stars, for which it is not.

\subsection{Mass Loss}

\begin{figure*}
\centering
\includegraphics[width=0.98\textwidth]{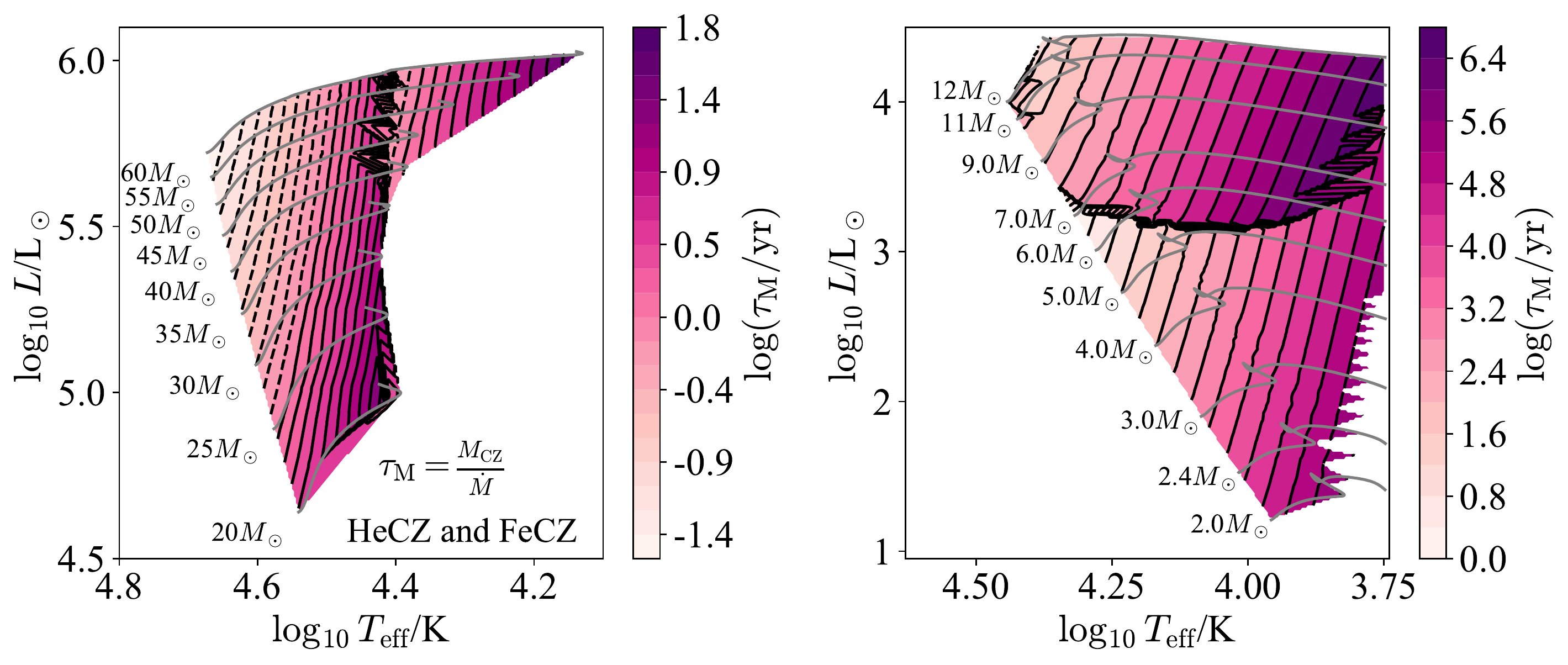}
\caption{The mass stripping time-scale given by equation~\eqref{eq:tauM} is shown on a Hertzsprung-Russel diagram in terms of $\log T_{\rm eff}$ and $\log L$ for stellar models ranging from $20-60 M_\odot$ (left) and $2-12 M_\odot$ (right) with $Z=0.02$. The sharp change at $\log L/L_\odot \approx 3.2$ is due to the appearance of the FeCZ. At low $\log T_{\rm eff}$ and low $\log L$ we have omitted regions where a vigorous H convection merges with the subsurface convection zones, and we do not report values calculated for the H convection zone in this diagram. Note that the two panels use different color scales.}
\label{fig:tauM}
\end{figure*}

The next time-scale of interest is the time for mass-loss to strip the region above the subsurface convection zone, given by
\begin{align}
	\tau_{\rm M} \equiv \frac{M_{\rm CZ}}{\dot{M}},
	\label{eq:tauM}
\end{align}
where $M_{\rm CZ}$ is the mass above the top of the convection zone and $\dot{M}$ is the mass-loss rate, evaluated using the prescription of~\citet{2001A&A...369..574V}.

Figure~\ref{fig:tauM} shows this time-scale.
Neglecting the sudden jump around $\log L/L_\odot \approx 3.2$, the contours of constant $\tau_{\rm M}$ are nearly vertical, meaning that this time-scale is very nearly a function of $T_{\rm eff}$ alone.
This is because the subsurface convection zones are caused by ionization features in the opacity, which are primarily a function of temperature.
This means that each convection zone is located at a location of nearly constant temperature in the star, so as $T_{\rm eff}$ varies the depth of the convection zone varies to hold the local temperature fixed.
As a result this time-scale very similar for stars of equal $T_{\rm eff}$ but very different $L$.
The jump at high luminosities arises because of a switch in which convection zone dominates, from the HeCZ to the FeCZ.

Beyond its scaling, $\tau_{\rm M}$ is notable for being so short: it is everywhere shorter than the main-sequence lifetime of the star.
This is particularly stark at the Zero-Age Main-Sequence, where the main-sequence lifetime is typically more than $10^5$ times longer than the mass-stripping time-scale.
As a result the entire layer above the subsurface convection zone is stripped away many times over, causing the convection zone to continuously move downward in mass coordinate in order to maintain a nearly-fixed temperature and hence depth from the surface.

\begin{figure*}
\centering
\includegraphics[width=0.98\textwidth]{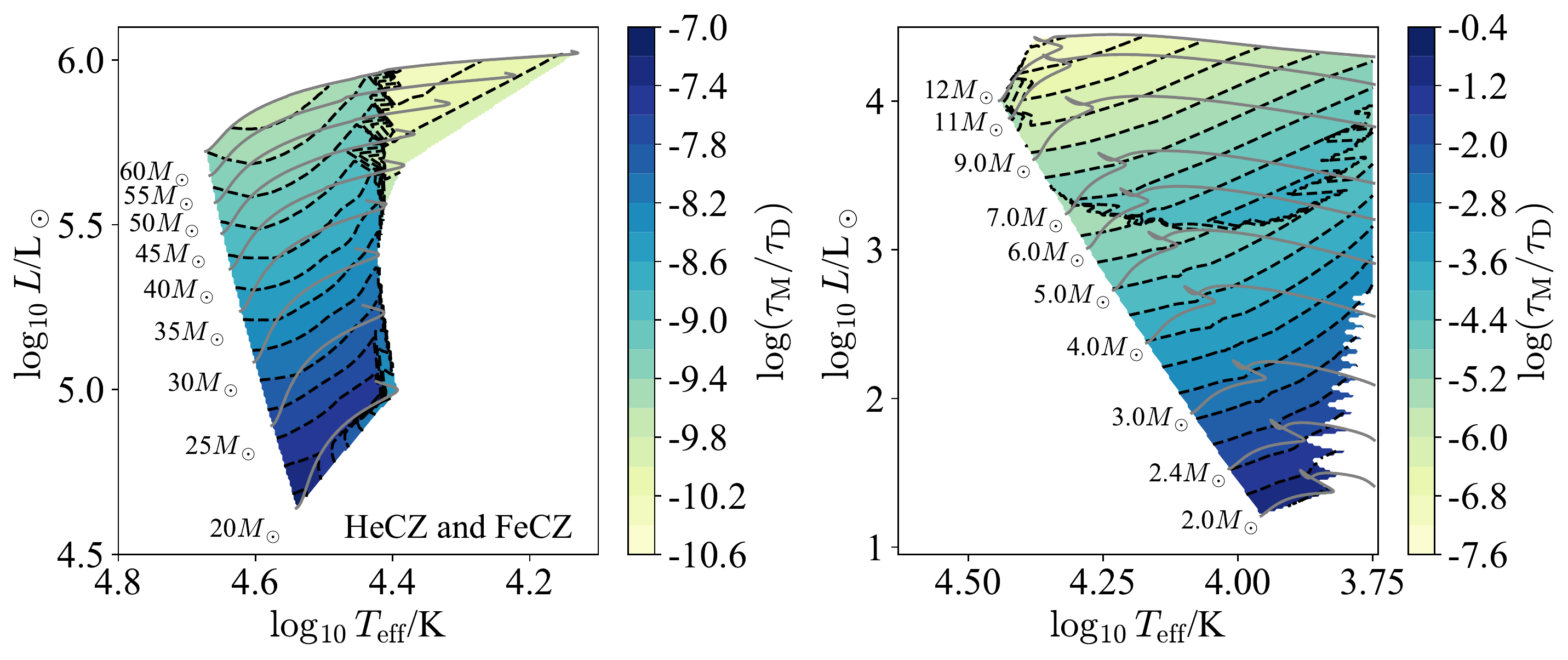}
\caption{The ratio of the mass stripping time-scale to the magnetic diffusion time-scale is shown on a Hertzsprung-Russel diagram in terms of $\log T_{\rm eff}$ and $\log L$ for stellar models ranging from $20-60 M_\odot$ (left) and $2-12 M_\odot$ (right) with $Z=0.02$. The sharp change at $\log L/L_\odot \approx 3.2$ is due to the appearance of the FeCZ. At low $\log T_{\rm eff}$ and low $\log L$ we have omitted regions where a vigorous H convection merges with the subsurface convection zones, and we do not report values calculated for the H convection zone in this diagram. Note that the two panels use different color scales.}
\label{fig:ratio}
\end{figure*}

In fact $\tau_{\rm M}$ is \emph{also} shorter than the magnetic diffusion time $\tau_{\rm D}$.
This may be seen in Figure~\ref{fig:ratio}, which shows the ratio $\tau_{\rm M}/\tau_{\rm D}$.

\subsection{Convective Turnover}

\begin{figure*}
\centering
\includegraphics[width=0.98\textwidth]{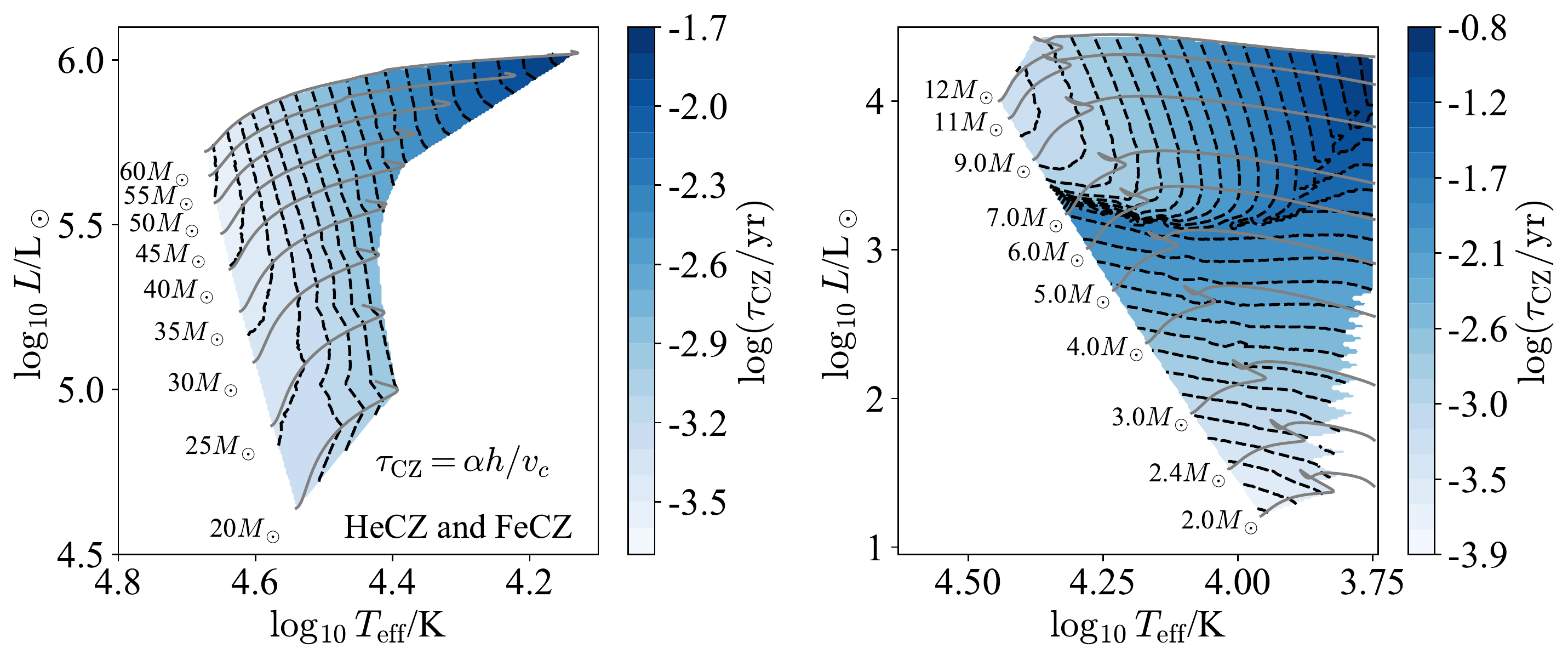}
\caption{The convective turnover time is shown on a Hertzsprung-Russel diagram in terms of $\log T_{\rm eff}$ and $\log L$ for stellar models ranging from $20-60 M_\odot$ (left) and $2-12 M_\odot$ (right) with $Z=0.02$. The sharp change at $\log L/L_\odot \approx 3.2$ is due to the appearance of the FeCZ. At low $\log T_{\rm eff}$ and low $\log L$ we have omitted regions where a vigorous H convection merges with the subsurface convection zones, and we do not report values calculated for the H convection zone in this diagram. Note that the two panels use different color scales.}
\label{fig:turn}
\end{figure*}

The final time-scale of interest, and also the shortest, is the convective turnover time 
\begin{align}
	\tau_{\rm C} \equiv \frac{\alpha h}{v_{\rm c}},
\end{align}
where $\alpha$ is the mixing length parameter, $h$ is the mean pressure scale height in the convection zone, and $v_{\rm c}$ is the mean convection speed.
This time-scale, shown in Figure~\ref{fig:turn}, is typically of order days or weeks, and so is many orders of magnitude shorter than any of the other time-scales we consider here.

\section{Scenario}\label{sec:story}

\begin{figure}
\centering
\includegraphics[width=0.45\textwidth]{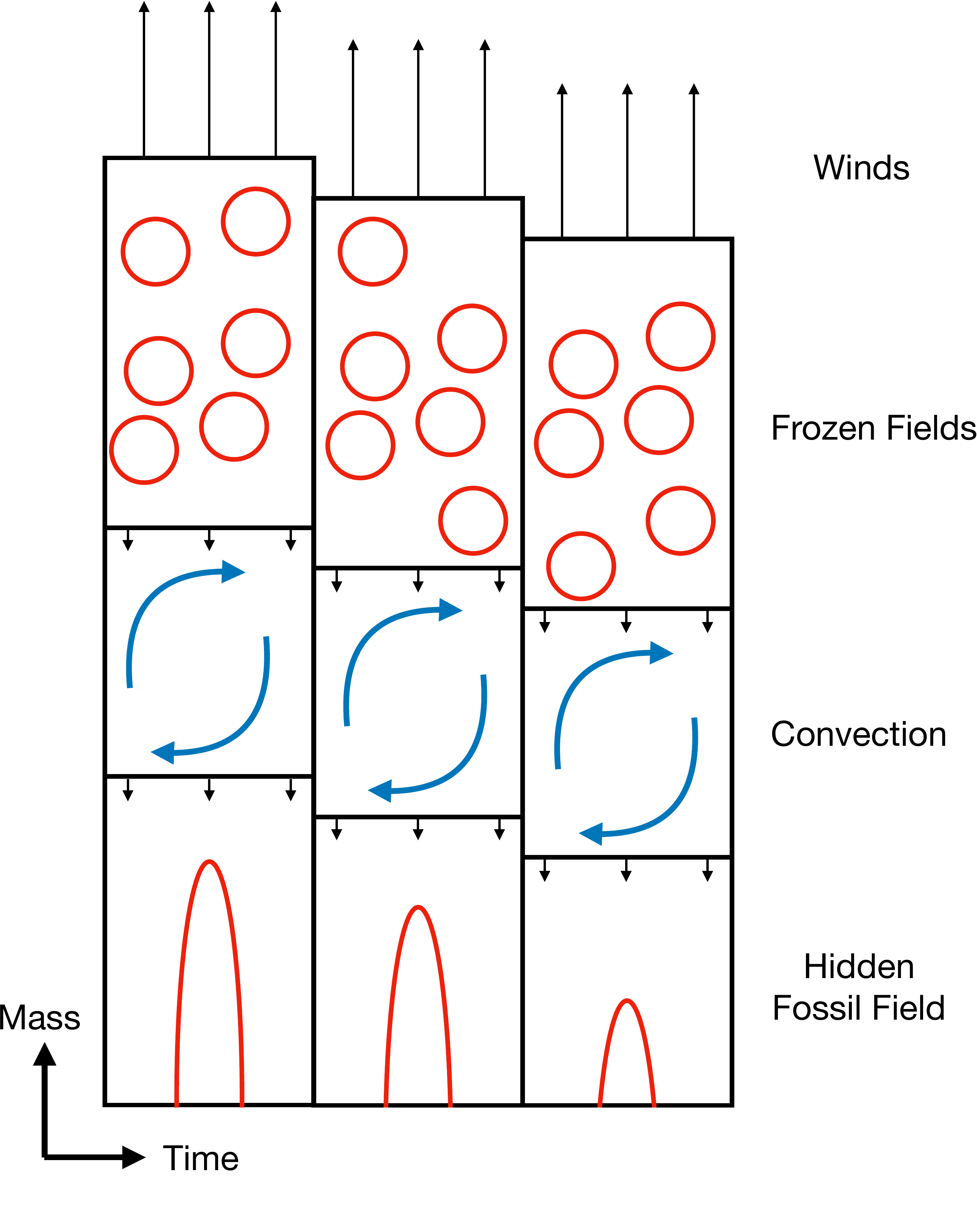}
\caption{The structure of the near-surface regions of an early-type star is shown schematically. The surface of the star is at the top, where mass loss results in a stellar wind. This causes the stellar surface to move downward in mass coordinate, and so the convection zone likewise moves downward, trailing a frozen-in layer of dynamo-generated magnetic fields. Beneath this is a fossil magnetic field hidden by this process. This fossil field could be quite strong, though it must be below the critical shutoff strength in order for convection to continue moving downward.}
\label{fig:schema1}
\end{figure}

Putting the various time-scales together, we have the following order
\begin{align}
	\tau_{\rm C} \ll \tau_{\rm M} \ll \tau_{\rm D}, \tau_{\rm MS},
	\label{eq:ordering}
\end{align}
where $\tau_{\rm MS}$ is the main-sequence stellar lifetime\footnote{In general $\tau_{\rm D}$ may be greater or less than $\tau_{\rm MS}$, but for our purposes all that matters is that both are much longer than the other two time-scales.}.
This ordering suggests how the near-surface magnetic field evolves in early-type stars with convective dynamos:
\begin{enumerate}
\item Mass loss strips material from the surface of the star.
\item Because the location of the convection zone is determined primarily by temperature, the convection zone moves downwards in mass coordinate to maintain an approximately constant depth below the surface.
\item Following~\citet{2020ApJ...900..113J}, we expect convection to eat away at any fossil field it encounters on the very short convective turnover time $\tau_{\rm C}$, reprocessing this into a disordered dynamo field of equipartition strength which it trails behind it.
\item Because the magnetic diffusion time is long compared with the mass-loss time, this dynamo-generated field is effectively frozen until mass-loss strips the surface down to it.
\item Mass-loss eventually reveals this frozen-in dynamo remnant and then removes it, revealing newer frozen fields beneath.
\end{enumerate}
The result is an evolution like that shown in Figure~\ref{fig:schema1}.

This scenario makes a number of predictions, which we discuss next.

\subsection{Field Loss}

In \citet{2020ApJ...900..113J} we discussed how a fossil field can be lost from the surface of an early type star: if the fossil field is weaker than the critical magnetic field required to shut off subsurface convection then convection proceeds to remove the near-surface component of the field. The hierarchy of timescales in~(\ref{eq:ordering}) shows that it should be difficult to catch a star in the act of losing its near-surface fossil field, because the relevant time-scale for that is at most the mass-stripping time $\tau_{\rm M}$, and more likely of order the convective turnover time, both of which are very short compared with the main-sequence stellar lifetime.

\subsection{Frozen Dynamo Fields}

Next, our scenario predicts that the subcritical magnetic fields observed at the surface of most early-type stars are the frozen-in remnants of a convective dynamo that passed through what is now the stellar surface.
This explains why those fields are observed to be remarkably stable on time-scales of months~\citep{2011A&A...532L..13P} and years~\citep{2010A&A...523A..41P}, which is not what would be expected for a field generated by a currently-active dynamo.

\subsection{Strong Hidden Fossil Fields}

A further prediction is that many early-type stars with subcritical surface magnetic fields may have strong fossil fields hiding just beneath their subsurface convection zones.
These fields would necessarily be subcritical~\citep{2020ApJ...900..113J}, but that still permits fields of order $300\,\mathrm{G}$ for A stars, $1\,\mathrm{kG}$ for B stars, and $3\,\mathrm{kG}$ for O stars~\citep{2020ApJ...900..113J}.
If these hidden fields evolve with flux conservation they could give rise to fields of order $10^6\,\mathrm{G}$ in Red Giant Cores/White Dwarfs, and $10^{15}\,\mathrm{G}$ in Neutron Stars.

This potentially resolves a number of puzzles.
First, astereoseismology reveals magnetic fields of order $10^{6}\,\mathrm{G}$ in the cores of Red Giants~\citep{2015Sci...350..423F}.
While these fields could be a result of the Tayler-Spruit dynamo, they are $10^{1-3}$ times stronger than dynamo models typically suggest~\citep{2002A&A...381..923S,2019MNRAS.485.3661F}, making hidden fossil fields an attractive alternative.
Another possible explanation is that the magnetic fields in Red Giant cores descend from main sequence convective core dynamos \citep{Cantiello:2016}.  

Secondly, it was recently noted that there are too-few strongly-magnetized massive stars for simple flux conservation to explain the observed magnetar population~\citep{2021MNRAS.tmp.1146M}.
A similar challenge exists at lower masses: $\sim 10\,\%$ of A/B star show strong surface magnetic fields~\citep{2007pms..conf...89P} while $\sim 20\,\%$ of their White Dwarf remnants have the $10^{5-6}\,\mathrm{G}$ fields those would produce under flux conservation~\citep{2019A&A...628A...1L}.
While both discrepancies could be evidence of late-stage evolutionary processes generating strong magnetic fields, it could also be that many massive stars with subcritical surface fields contain hidden strong fossil fields, which are subsequently revealed in their compact remnants.

\subsection{Cyclical Wind Variability}

Finally, this scenario may explain the coherence time of cyclical wind variability and Discrete Absorption Components (DACs) in more massive O stars ($20-50 M_\odot$).

The large majority of O-type stars, as well as many B stars, show unexplained cyclical variability in their spectral lines \citep{1996yCat..41160257K,1999A&A...344..231K}.
The most prominent feature of this variability are DACs, which appear as excess absorption features in the UV/optical-wind line profiles that migrate towards higher velocities with time~\citep{Kaper:1997}.
DACs are therefore believed to be due to overdense features propagating through and accelerating with the wind~\citep{1996ApJ...462..469C}.

Notably this variability appears as a modulation on the
rotational timescale, but is not strictly periodic \citep{Kaper:1994}.
DACs persist for several rotation periods, and have a typical lifetime of order $10$ days~\citep{1995ApJ...452L..53M,1996A&AS..116..257K,2014MNRAS.441..910R}. It also appears  that these features are seeded at --or close to-- the stellar surface \citep[e.g.][]{Kaper:1997}.  While their exact origin is currently unknown, two possible mechanisms for explaining DACs are discussed in the literature: surface magnetism \citep[e.g.][]{2011A&A...534A.140C,Sudnik:2016}  and non-radial pulsations  \citep[NRPs, e.g.][]{Lobel:2008} .

If these features are caused by surface magnetic fields, and if our story holds, then the lifetime of the DACs ought to correspond to the time over which mass loss strips away the surface magnetic field to reveal a deeper layer of the frozen-in field. For dynamo-generated fields we expect the magnetic field to vary over a depth of order a pressure scale height in the convection zone.
The FeCZ is the dominant convection zone in these stars, and one pressure scale height in the FeCZ is of order 50~per-cent of the distance from the top of the FeCZ to the surface of the star~\citep[see e.g. ][, Figure 1, for a $20 M_\odot$ model]{Cantiello_2021}.
So a time-scale of roughly $1/2$ of the mass-stripping time is what we expect.

Figure~\ref{fig:tauM} shows the latter time-scale for stellar models ranging from $20-60M_\odot$.
For much of the main-sequence, particularly at higher masses, $\tau_{\rm M}$ is of order $10-100\,\rm d$, giving a prediction of $5-50\,\mathrm{d}$ for the DAC lifetime, in reasonable agreement with the observer $10\,\mathrm{d}$ coherence time.
It seems plausible then that cyclical wind variability in O stars may be gradually revealing ever-deeper layers of a frozen-in dynamo-generated magnetic field.

If this is correct then we should expect the coherence times of DACs and cyclical wind variability to depend strongly on mass and age along the main-sequence.
Towards the end of the main-sequence and towards lower masses ($10-20 M_\odot$) we should expect coherence times of order years, while the time-scale should shorten to just a few days for the most massive stars early on the main-sequence, and indeed may be sufficiently short to not be identifiable as a periodic signal corotating with the stellar surface.
We are not aware of observations which probe this prediction, and this seems like a promising avenue to test our theory.
 
\section{Limitations}\label{sec:limits}

While our scenario potentially resolves a number of observational puzzles, there are several challenges to it.

First, while there have been numerical simulations of convection processing a pre-existing magnetic field~\citep{2001ApJ...549.1183T,Featherstone:2009,2021MNRAS.503..362K}, we are not aware of any that have studied the extreme aspect ratios ($\sim 10^2$) relevant for subsurface convection zones.
The aspect ratio is potentially important because, in order to cause reconnection in large-scale fossil fields, small-scale convective motion has to twist the field over large scales comparable to the stellar radius.
Given time we suspect that this occurs, but we have no proof of it nor are we aware of simulations or rigorous arguments demonstrating it.
Testing this hypothesis is therefore crucial for understanding our theory.

\begin{figure*}
\centering
\includegraphics[width=0.98\textwidth]{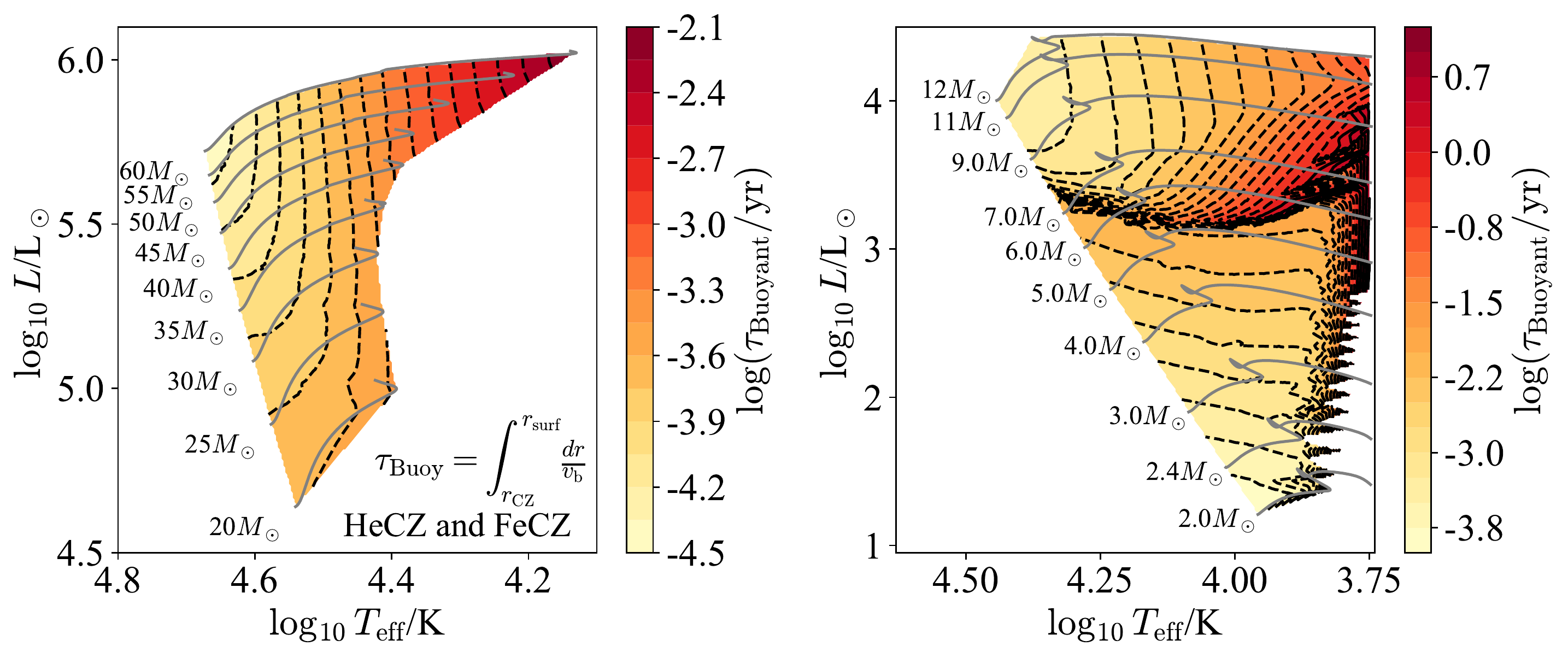}
\caption{The buoyancy time-scale $\tau_{\rm Buoy}$ is shown on a Hertzsprung-Russel diagram in terms of $\log T_{\rm eff}$ and $\log L$ for stellar models ranging from $20-60 M_\odot$ (left) and $2-12 M_\odot$ (right) with $Z=0.02$. The sharp change at $\log L/L_\odot \approx 3.2$ is due to the appearance of the FeCZ. At low $\log T_{\rm eff}$ and low $\log L$ we have omitted regions where a vigorous H convection merges with the subsurface convection zones, and we do not report values calculated for the H convection zone in this diagram. Red dots show the location of 20\% increase in fractional age, from the zero age main sequence to hydrogen exhaustion. Note that the two panels use different color scales.}
\label{fig:buoy}
\end{figure*}

Secondly, there is an additional time-scale we have not considered, namely the that for detached magnetic field loops in the radiative zone to rise buoyantly.
This scale was investigated by~\citet{2011A&A...534A.140C}, who found it to be very short.
Following their prescriptions we calculated the time for a field loop of radius one pressure scale height to rise from the dominant convection zone to the surface of the star, shown in Figure~\ref{fig:buoy}.
This scale is much shorter than either the magnetic diffusion or mass-loss time-scales, typically of order hours, and so if buoyancy plays a significant role it could change our story fundamentally.

In order for the buoyancy time-scale to matter it must be that the magnetic field indeed forms detached field loops, and that the typical magnetic field strength in the radiative zone is less than that of these loops, so that they rise.
Both of these are open questions, and a wide range of possibilities seem plausible.
For instance one option is that small detached field loops could rise and fall depending on their amplitudes amidst a larger-scale frozen and stable background field, in which case it could be that there is variation on the buoyancy time-scale but that that variation is not visible, because the larger-scale field is what dominates the observations.
This, too, seems like an area where numerical simulations will likely play a key role.

To the extent that there is observational evidence regarding the importance of buoyancy, it appears to be against.
The magnetic fields of A/B stars are observed to be stable over time-scales of months~\citep{2010A&A...523A..41P,2011A&A...532L..13P}, which is much longer than the hour-scale buoyancy time these stars.
Likewise if buoyancy really drives surface magnetic variability we should expect DACs to have coherence times of the same order, which for massive O stars is minutes to hours (Figure~\ref{fig:buoy}), whereas the observed coherence times are of order $10$ days.
Both of these lines of reasoning suggest that magnetic buoyancy does not play a significant role in setting the evolution of the large-scale observable surface magnetic field, though it could plausibly still matter for determining the field structure at smaller scales.

Next, the mass stripping timescales derived in Figures~\ref{fig:tauM} and \ref{fig:ratio} were calculated assuming the \citet{2001A&A...369..574V} mass loss rate.
Plots of $\dot{M}$ across the Hertzsprung-Russel diagram are provided in Appendix~\ref{appen:mdot}.
This recipe accounts for line-driven winds, which are fairly well understood from both a theoretical and an observational standpoint. 
This said, some uncertainties remain due to the existence of wind clumping, which could cause a reduction of the mass loss rate by a factor of up to $\approx3$  \citep[see e.g.][]{Smith:2014}. Moreover, there are some indications that line driving is inefficient in later O-type stars (hot dwarfs): their observed mass loss rates seems to be much lower than the theoretical predictions\footnote{Though see e.g.~\citep{2021A&A...648A..36B} for predictions which agree better with observations.}, the so-called ``weak-wind problem'' \citep{Puls:2008,Muijres:2012}.
Additionally there are predictions that strong magnetic fields may further reduce mass loss~\citep{2008MNRAS.385...97U,2017MNRAS.466.1052P}.
Finally, the stellar winds of A and late B-type stars are difficult to study observationally. This is expected, since the predicted  mass loss rates are very low \citep{Krticka:2014}, and larger uncertainties should be considered for our predictions in this regime.  

With these uncertainties in mind, for early-type stars our predictions should probably be considered lower limits for the mass stripping timescale.
While most of our scenario is robust to this uncertainty, if the true mass loss rates are more than a factor of several less than those we have assumed then we no longer find quantitative agreement with the DAC time-scales.
In that event it is possible to still fit the observations by e.g. assuming that the characteristic length-scale for the magnetic field is shorter than a pressure scale-height, but we have no particular physical basis for doing so and such an argument would likely require motivation from simulations.

Finally, an important challenge to our scenario is the observation of ordered (large scale) yet sub-critical magnetic fields in O/B stars.
For instance, all but one of the O-stars in the sample of~\citet{2013MNRAS.433.2497S} had sub-critical magnetic fields and showed macroturbulence, consistent with our story, yet several of these also appear to be stable over time and were consistent with strong dipolar fields~\citep[e.g.][]{2010MNRAS.407.1423M,10.1093/mnras/stv2568}.
That these fields are stable is not a particular challenge: for most objects the observational baseline is at most a decade, which is comparable to or smaller than the mass-loss time-scale (Figure~\ref{fig:tauM}) and so this apparent stability is actually consistent with our predictions.
More challenging, however, is the fact that these magnetic fields have a large dipolar component, indicating large-scale structure.

One way to reconcile these observations with our theory is to note that dynamos can generate strong dipolar components.
This is seen in the Sun, where there is a dominant dipolar component along with a decaying spectrum of higher order moments~\citet{2019A&A...623A..51Z}.
It is not inconceivable that a similar process could be at work in a small fraction of early-type stars, though there are theoretical expectations that the field become more ordered the lower the convective Rossby number~\citep{10.1111/j.1365-246X.2006.03009.x}, and we expect early-type stars to have much greater Rossby numbers than the Sun because their turnover time-scales are 1-2 orders of magnitude shorter than the solar value of $\sim 10-30\,\mathrm{d}$.

To the extent possible, then, it seems valuable to pick a small number of these objects and investigate them in detail.
In particular, it would be useful to characterize the power spectrum of the magnetic field as precisely as possible to better understand if the field is purely dipolar or if there are significant but subdominant higher order moments.
If these truly are pure sub-critical dipolar fields then some part of our story must break down, and it would be very interesting to understand in more detail the properties of these stars (mass, age, rotation, etc.).

Along the same lines, it would be helpful to look for trends of dipolar character with rotation.
If these magnetic fields are dynamo-generated as we suggest they should show a preference to be more dipolar with increasing rotation rate, whereas we are not aware of a reason to expect such a correlation with fossil fields.

\section{Discussion}\label{sec:discussion}

We have proposed a model in which subsurface convection zones move downward in mass coordinate as stars lose mass.
Because magnetic diffusion is slow compared with mass loss, this results in the convection zone trailing a dynamo-strength frozen-in magnetic field above it.
The key predictions of this model are:
\begin{enumerate}
\item Stars lose their surface fossil fields quickly, so magnetic field measurements are either of fossil fields or dynamo-generated fields, not a transition state.
\item When massive stars have subcritical surface magnetic fields, those fields are the frozen-in remnants of a convective dynamo.
\item Because subcritical surface fields are frozen-in, they vary on the mass-loss time-scale rather than the much faster convective turnover time. This potentially explains the coherence time of cyclical wind variability and DACs.
\item Massive stars with subcritical surface magnetic fields may have strong internal fossil fields, up to the convective shutoff strength~\citep{2020ApJ...900..113J}.
\item These strong internal fields may explain the apparent overabundance of strongly-magnetized White Dwarfs and Magnetars without any further field generation.
\end{enumerate}
These predictions, however, are contingent on a number of assumptions about how convection and magnetic fields interact.
In particular, we have assumed that:
\begin{enumerate}
\item Convection can readily process strong sub-critical magnetic fields into weaker equipartition dynamo fields, even at high aspect ratios.
\item The resulting frozen-in fields do not buoyantly rise.
\end{enumerate}
These assumptions are important for matching the observations, but have not been verified by any rigorous calculation or simulation.
Testing them is therefore an important challenge for the future.

Our scenario might also be relevant for understanding the details of chemical anomalies formation.
Diffusion processes in A and B stars hosting strong fossil fields may be responsible for the appearance of chemical peculiarities.
If the atmosphere is stabilized by magnetic fields, gravitational settling and radiative levitation become important \citep{Michaud:1970,Michaud:1976,Richer:2000}

Some abundance anomalies are also observed in some stars with subcritical magnetic fields like Sirius B. Theoretical calculations can reproduce these anomalies by assuming that either turbulent mixing or wind mass loss compete with atomic diffusion processes \citep{Michaud:2011}. In the context of this work, stars with supercritical magnetic fields do not possess near surface convective layers and diffusion processes can operate relatively undisturbed. On the other hand, stars with subcritical magnetic fields are able to efficiently stir layers of material that are subsequently advected to the stellar surface, which likely results in much smaller surface abundance anomalies. 
\acknowledgments

The Flatiron Institute is supported by the Simons Foundation. 
\software{
\texttt{MESA} \citep[][\url{http://mesa.sourceforge.net}]{Paxton2011,Paxton2013,Paxton2015,Paxton2018,Paxton2019},
\texttt{MESASDK} 20190830 \citep{mesasdk_linux,mesasdk_macos},
\texttt{matplotlib} \citep{hunter_2007_aa}, 
\texttt{NumPy} \citep{der_walt_2011_aa}
         }

\clearpage

\appendix

\section{MESA} \label{appen:mesa}

Calculations were done with revision 15140 of the Modules for Experiments in Stellar Astrophysics
\citep[MESA][]{Paxton2011, Paxton2013, Paxton2015, Paxton2018, Paxton2019} software instrument.

The MESA EOS is a blend of the OPAL \citep{Rogers2002}, SCVH
\citep{Saumon1995}, FreeEOS \citep{Irwin2004}, HELM \citep{Timmes2000},
and PC \citep{Potekhin2010} EOSes.

Mixing Length Theory was implemented following the prescription of~\citet{1968pss..book.....C} using a mixing length parameter $\alpha=1.6$ 

Radiative opacities are primarily from OPAL \citep{Iglesias1993,
Iglesias1996}, with low-temperature data from \citet{Ferguson2005}
and the high-temperature, Compton-scattering dominated regime by
\citet{Buchler1976}.  Electron conduction opacities are from
\citet{Cassisi2007}.

Nuclear reaction rates are from JINA REACLIB \citep{Cyburt2010} plus
additional tabulated weak reaction rates \citet{Fuller1985, Oda1994,
Langanke2000}.
Screening is included via the prescription of \citet{Chugunov2007}.
Thermal neutrino loss rates are from \citet{Itoh1996}.

The inlists used to run our models are provided by~\citet{adam_s_jermyn_2021_5076469}. 
\section{Mass Loss}\label{appen:mdot}

\begin{figure*}
\centering
\includegraphics[width=0.98\textwidth]{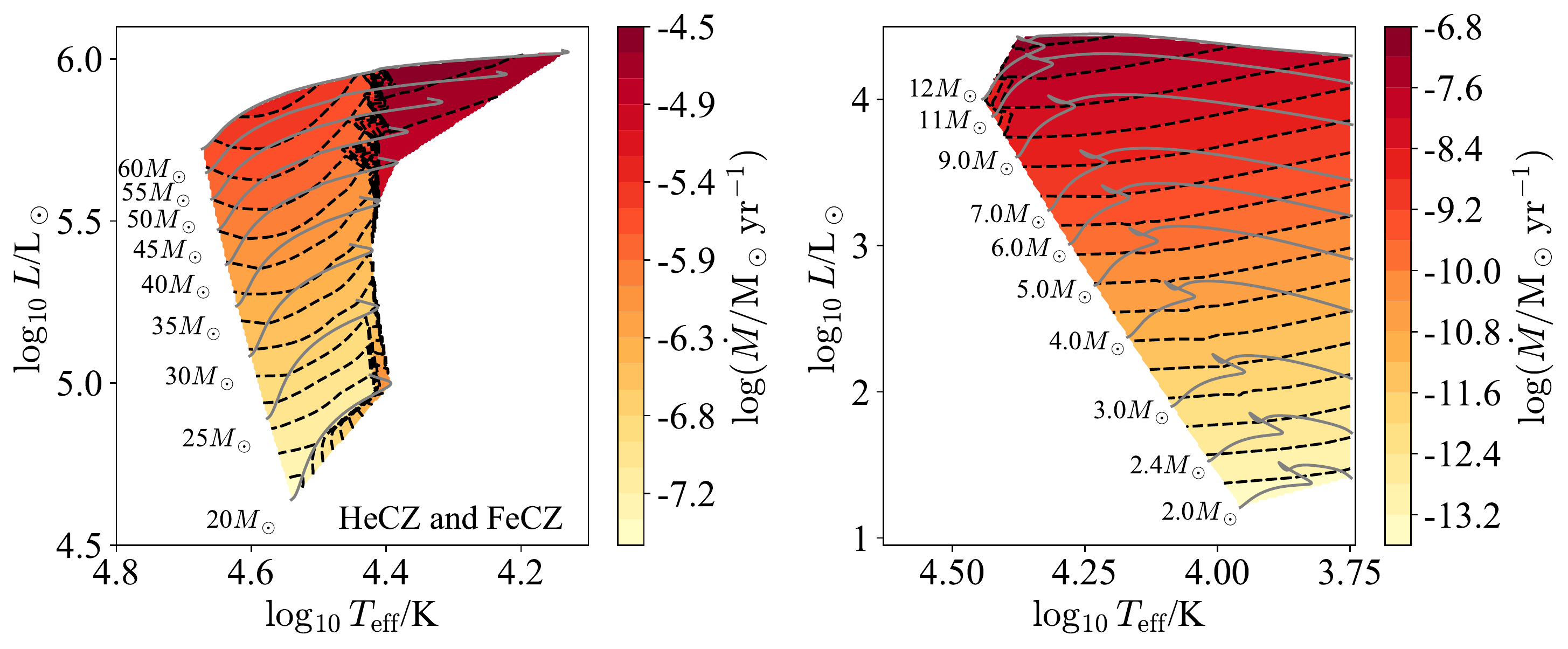}
\caption{The mass loss rate is shown on a Hertzsprung-Russel diagram in terms of $\log T_{\rm eff}$ and $\log L$ for stellar models ranging from $20-60 M_\odot$ (left) and $2-12 M_\odot$ (right) with $Z=0.02$.  Note that the two panels use different color scales. The blue point in the lower-right corresponds to Sirius A, with parameters corresponding to those of~\citet{2011A&A...534A..18M}.}
\label{fig:mdot}
\end{figure*}

Figure~\ref{fig:mdot} shows the mass loss rate given by~\citet{2001A&A...369..574V} for stellar models up to $60 M_\odot$.
Note the bistability feature around $\log T/\mathrm{K} \approx 4.4$.

\bibliography{refs}
\bibliographystyle{aasjournal}

\end{document}